\pdfoutput=1
\documentclass{PoS}

\usepackage{amsmath,amssymb}
\usepackage{subfigure}

\def\bx{\bf x}
\def\bxp{\bf x\prime}
\def\by{\bf y}

\def\bp{\bf p}

\title{Investigations of QCD at non-zero isospin density}

\ShortTitle{Investigations of QCD at non-zero isospin density}

\author{\speaker{Zhifeng Shi}\\
       Department of Physics, College of William and Mary, Williamsburg, VA 23187, USA\\
       Jefferson Laboratory, 12000 Jefferson Avenue, Newport News, VA 23606, USA \\
       E-mail: \email{zshi@email.wm.edu}}

\author{{William Detmold} \\
       Department of Physics, College of William and Mary, Williamsburg, VA 23187, USA\\
       Jefferson Laboratory, 12000 Jefferson Avenue, Newport News, VA 23606, USA \\
       E-mail: \email{wdetmold@wm.edu}}

\abstract{ We investigate the QCD phase diagram 
  as a function of isospin chemical potential at a fixed
  temperature by directly putting large numbers of $\pi^+$s
  into the system. Correlation functions of \mbox{$N$ $\pi^+$-systems}
   involves $N!N!$ contractions, and become 
  extremely expensive when $N$ is large.
   In order to alleviate this problem, a recursion relation
  of correlation functions has been derived in Ref.~\cite{Detmold:Savage}
  that substantially reduces the number of independent contractions
  needed and makes the study of many pions systems be 
  possible. In this proceeding this method is investigated numerically.
  We have also constructed a new method 
  that is even more efficient, enabling us to study systems of up to \mbox{$72$ $\pi^+$'s}.
 }

\FullConference{XXIX International Symposium on Lattice Field Theory \\
                July 10-16 2011\\
                Squaw Valley, Lake Tahoe, California}

\begin{document}

\section{Introduction}
 The QCD phase diagram has been studied with many methods,
 but there are still many questions waiting 
 to be answered.  At zero isospin density,
 there is a crossover (phase transition)
 from a confinement phase at low temperature to 
 a decomfinement phase at higher temperature. From 
 chiral perturbation theory  ($\chi$PT), 
 we expect that when the isospin chemical potential
 reaches the mass of a single pion, pions start to condense,
 and at higher isospin chemical potential there might be
 a crossover or a phase transition to another state~\cite{Son:Stephanov}. 
 Systems containing up
 to $12$ $\pi^+$s have been studied
 by directly computing the required contractions in
 Ref.~\cite{Detmold:Savage:Torok} \cite{Silas:Will},
 finding agreement with the expectations from $\chi$PT. In order to
 put more $\pi^+$s into the system,
 a second source is required because of the Pauli principle, and studying systems
 of more than $12$ $\pi^+$ becomes almost impossible by directly computing
 all possible contractions.

 In Ref.~\cite{Detmold:Savage}, a recursion relation of correlation functions
 of $n$-$\pi^+$ systems is constructed which 
 significantly reduces the amount 
 of work needed, enabling us to 
 study systems up to $24$ $\pi^+$'s. Recursion relations 
 of $M$ species systems have also been discussed
 in the same paper. Even with the recursion relation, 
 studying a 3-source system (36 pions) still consumes
 a substantial amount of time. In order to overcome 
 this problem we have constructed a new method which
 allows practical calculation of up
 to $72 \pi^+$s.

 In this proceeding, we first 
 review the methodology of the recursion relation in
 position space and then extend its application to momentum space,
 after that we present studies of systems containing up 
 to $24 \pi^+$s by applying the recursion relation in 
 momentum space. At last we present results from
 the new method to study large number of pion systems. 

 \section{Methodology of the recursion relation}
 \subsection{Recursion relation in position space}
   In order to explore systems containing
   up to $12M$ $\pi^+$'s, M different sources
   are required because of the Pauli
   principle. A correlation function of
   a system with $n_{i}$-$\pi^+$s in the $i^{th}$ source
   is:
\begin{eqnarray}
  C_{( n_1\pi^+_1\ , ... , \ n_m \pi^+_m )}(t) & = &
  \left \langle\
  \left(\ \sum_{\bf x}\ \pi^+({\bf x},t)\ \right)^{\overline{n}}
  \left( \phantom{\sum_{\bf x}}\hskip -0.2in
    \pi^-({\bf  y_1},0)\ \right)^{n_1} ...
  \left( \phantom{\sum_{\bf x}}\hskip -0.2in
    \pi^-({\bf  y_m},0)\ \right)^{n_m}\
  \right \rangle
  \ ,
\end{eqnarray}
 where ${\bar n} = \sum_{i=1}^m n_i$. 
 Calculating this correlation function according to
 Wick's theorem involves ${\bar n}!{\bar n}!$ contractions,
 which make the study for a system of large number
 of $\pi^+$'s extremely time consuming. However
 the recursion relation of uncontracted correlation functions
 $Q_{(n_1 , n_2 , ... , n_m) }(t)$,
 makes
 the study of such systems feasible. The correlation function
 can be expanded as:
  \begin{eqnarray}
   C_{( n_1\pi^+_1\ , ... , \ n_m \pi^+_m )}(t) \ =\
   (-)^{\overline{n}}\
  \left(\ \prod_i\ n_i! \ \right)\
  \langle\ Q_{(n_1 , n_2 , ... , n_m) }(t)\ \rangle
  \ \ \ \ ,
   \end{eqnarray}
 where$\langle \ \rangle$ denotes spin color trace,
  and $Q_{(n_1 , n_2 , ... , n_m) }(t)$
 satisfies the following ascending recursion
 relation:
\begin{eqnarray}
  Q_{(n_1+1 , n_2 , ... , n_m)}& = &
  \langle\ Q_{(n_1 , n_2 , ... , n_m)} \rangle\ P_1
  \ -\
  \overline{n}\ Q_{(n_1 , n_2 , ... , n_m)} P_1
  \nonumber\\
  & ... & +\
  \langle\ Q_{(n_1+1 , n_2 , ... n_k-1, ... ,  n_m)} \rangle\ P_k
  \ -\
  \overline{n}\ Q_{(n_1+1 , n_2 , ... n_k-1 , ... , n_m)} P_k
  \nonumber\\
  & ... & +\
  \langle\ Q_{(n_1+1 , n_2 , ... , n_m-1)} \rangle\ P_m
  \ -\
  \overline{n}\ Q_{(n_1+1 , n_2 , ... , n_m-1)} P_m
  \ \ \ \ ,
\end{eqnarray}
 and the initial conditions are
 $Q_{(1,0,...,0)} = P_1 = A_1,Q_{(0,1,...,0)} = P_2 = A_2, \cdots $,
 where $A_i$'s are uncontracted single pion correlators.
Descending recursion relations
can also be derived:
\begin{eqnarray}
  Q_{\bf n} = \sum_{k=1}^M\frac{1}{N+1-{\bar n}}\langle Q_{{\bf n}+{\bf 1}_k} A^{-1}
                \left( P_k\cdot A^{-1}\right ) \rangle \cdot I_N
                - Q_{{\bf n}+{\bf 1}_k}A^{-1}\left( P_k\cdot A^{-1}\right)
\end{eqnarray}
 where ${\bf n} = (n_1,n_2,\cdots, n_m)$,
 ${\bf 1}_k = (0,0,\cdots,1,0,\cdots)$ with only
 the $k^{th}$ nonvanishing unit element,
 and
 $Q_{12,\ldots,12}$, 
 $P_k$ and $A$ are constructed in the following way:
\begin{eqnarray}
 Q_{12,\ldots,12} &=& (N-1)!det(A)\cdot I_{N} \nonumber\\
  P_k  &=&
  \left(\begin{array}{c|c|c|c}
      0 & 0 & 0 & 0 \\
      \hline
      \vdots & \ldots & \ldots & \ldots \\
      \hline
      A_{k1}(t)&A_{k2}(t) & \ldots & A_{kM}\\
      \hline
      \vdots & \ldots & \ldots & \ldots \\
      \hline
      0 & 0 & 0 & 0
    \end{array}
  \right)
  \ \, \ \
  A  =
  \left(\begin{array}{c|c|c|c}
      A_{11}(t)&A_{12}(t) & \ldots & A_{1M}\\
      \hline
      \vdots & \ldots & \ldots & \ldots \\
      \hline
      A_{k1}(t)&A_{k2}(t) & \ldots & A_{kM}\\
      \hline
      \vdots & \ldots & \ldots & \ldots \\
      \hline
      A_{M1}(t)&A_{M2}(t) & \ldots & A_{MM}\\
    \end{array}
  \right)
\end{eqnarray}
 where $A_{i,j}(t)$ is defined as
 $A_{i,j}\left(t\right) = \sum_{\bf x} S\left({\bf x}_i,{\bf x}\right)S^+\left({\bf x}_j,{\bf x}\right)$.
 
 One way to construct $A_{i,j}$ is shown in the
 left plot of Fig.~\ref{fig:construct_uncorr}. 
 Correlation functions of two species from
 multiple sources have similar recursion relations,
 for detail discussion about 
 the recursion relations see Ref.~\cite{Detmold:Savage}.
  \subsection{The recursion relation in momentum space}
\begin{figure}
   \centering
   \subfiguretopcapfalse
   \subfigure[]{\includegraphics[width=6.0cm]{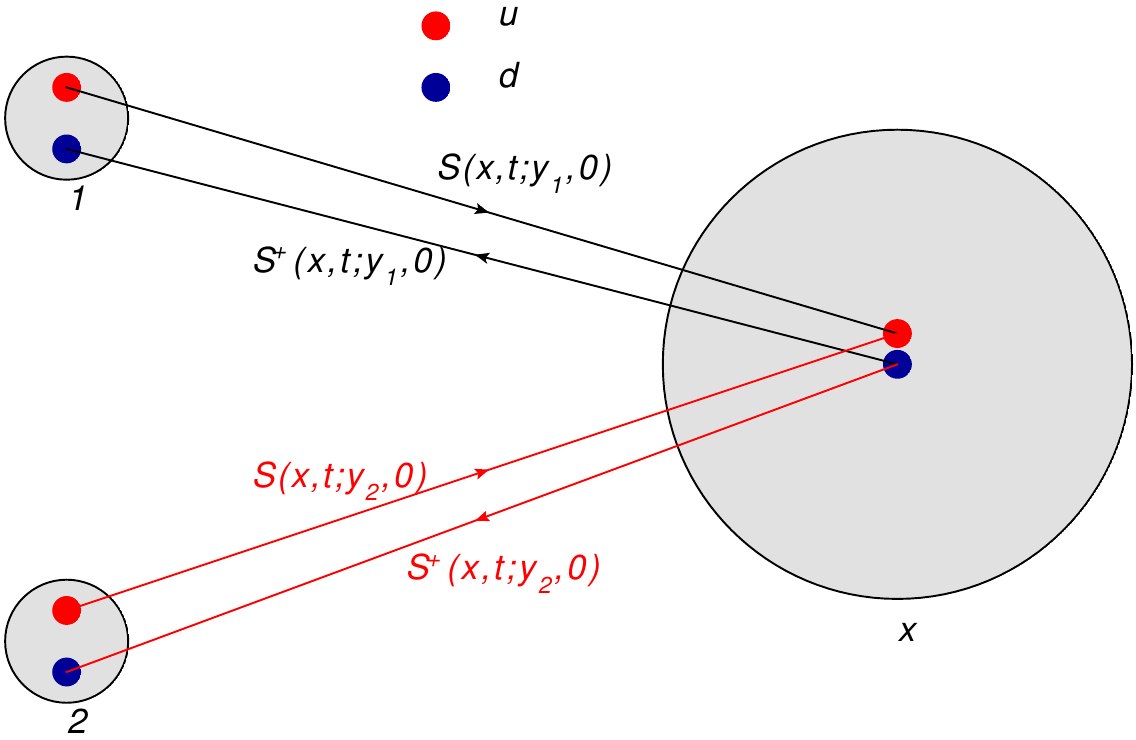}}
   \hspace{1.0cm}
    \subfigure[]{\includegraphics[width=6.0cm]{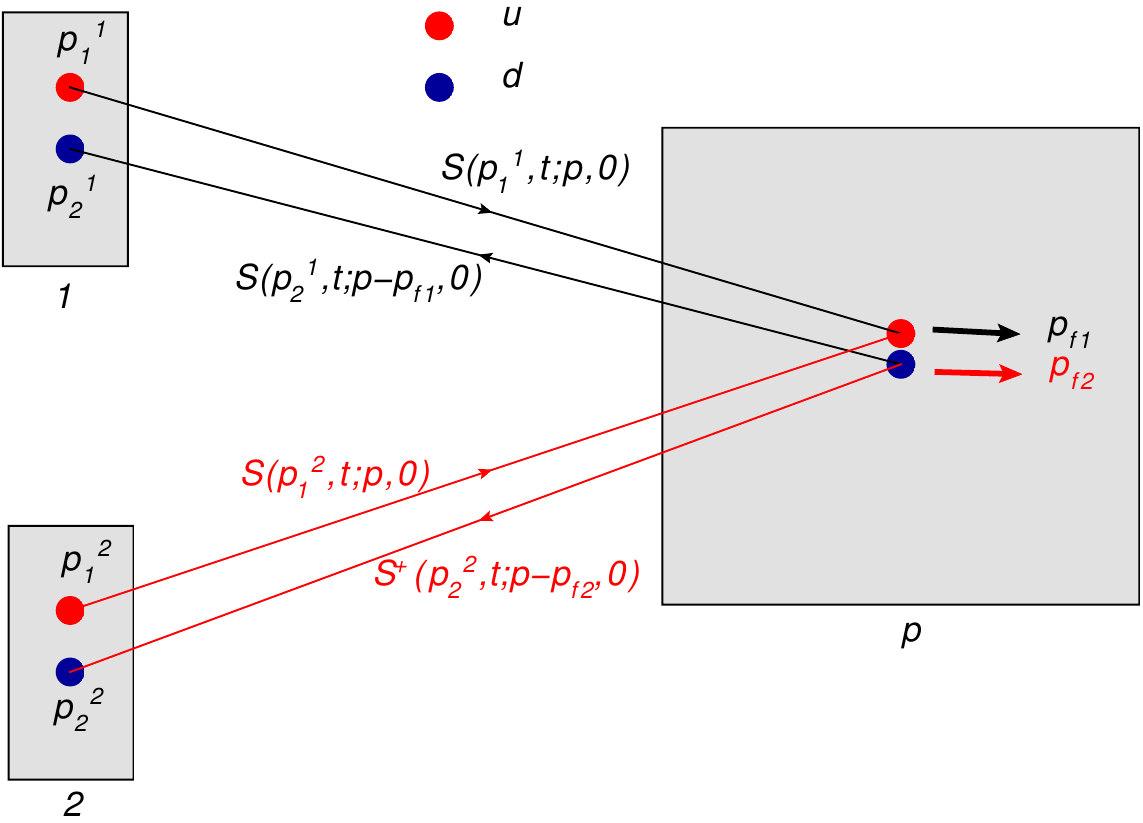}} \\
   \caption{The left figure shows how to construct uncontracted correlators in 
    spatial space and the right figure shows how to get its counterparts in
    momentum space. $A_{i,j}$ is constructed by following the line from source $i$ 
    to $x(p)$, returning to source $j$, multiplying 
    the $S(S^{\dagger})$ with respect to each line and summing over $x(p)$.}
   \label{fig:construct_uncorr}
   \end{figure}

     A correlation function of a system
     having $n_1$-$\pi^+$s
     in the first source and $n_2$-$\pi^+$s in another source
     with total momentum $n_1{\bp}_{f_1}+n_2{\bp}_{f_2}$
     is:
     \begin{eqnarray}
        C_{n_1\pi^+,n_2\pi^+}\left(t\right) = \left \langle \prod_{i=1}^2\left( \sum_{{\bx}_i,{\bxp}_i}
                       e^{-i\left({{\bp}^i_1 {\bx}_i-{\bp}^i_2 {\bxp}_i}\right)}
                {\overline u}\left({\bx}_i,t\right)\gamma_5 d\left({\bxp}_i ,t\right)\right)^{n_i}
                \cdot
                \prod_{j=1}^{\overline n}\left(\sum_{{\by}_j} e^{i{\bp}_{f_j} {\by}_j}
                  {\overline d}\left({\by}_j,0\right)\gamma_5
                           u\left({\by}_j,0\right)\right) \right \rangle \nonumber
     \end{eqnarray}
     where ${\overline n} = n_1+n_2$.
     Momentum conservation requires that
    $n_1{\bp}^1_1+n_2{\bp}^2_1 - n_1{\bp}^1_2 - n_2{\bp}^2_2 = \sum_{j=1}^{\overline n}{\bp}_{f_j}$
    must be satisfied to get non-vanishing $C_{n_1\pi^+,n_2\pi^+}$.
     Each choice of
     ${\bp}^i_j, i,j=1,2$ satisfying this relation is a separate measurement.
     By replacing propagators in position space
     by propagators in momentum
     space, a similar
     recursion relation still holds. The only
     difference is the construction of
     uncontracted correlation functions $A_{i,j}$ defined as
     $A_{i,j}\left(t\right) = \sum_{\bp} S\left({\bp}^1_i,{\bp}\right)S^+\left({\bp}^2_j,{\bp}-{\bp}_{f_j}\right)$,
     which are compared on Fig.~\ref{fig:construct_uncorr}.

 \section{Results}

   Because of the finiteness of the
   temporal extent and the factorisable
   nature of the multi-hadron systems being
   constructed, thermal effects are particularly
   important in multi-hadron systems and the correlation functions assume
    the form: 
  \begin{eqnarray}
   \label{eqn:corr_expected}
   C_{n\pi^+}\left(t\right) &=& \sum_{m=0}^n {n \choose m} Z_m^n e^{-\left(E_m+E_{n-m}\right)T/2}
             \cosh\left(\left(E_m-E_{n-m}\right)\cdot \left(t-T/2\right)\right)+\cdots 
  \end{eqnarray}
  where dots represent higher excitations, $T$ is
  the temporal extent and $E_n$ is the energy of a rest system of $n$-$\pi^+$s.
  The ground state comes from all $\pi^+$'s propagating in the
  same direction in time, and thermal states are from some
  $\pi^+$'s propagating in one direction while the rest propagate
  in the opposite direction.

  \subsection{Verify the dispersion relationship}

  Calculations have been performed using
  anisotropic $\{16^3, 20^3, 24^3\} \times 128$ lattices
  with anisotropy
  \mbox{$\xi=3.5$} at a quark mass corresponding to a $m_{\pi}=390$Mev. On the lattice,
  only discrete momentum $\frac{2\pi}{L}n$
  are allowed.
  $E_{m\pi^+}$ of systems with total momentum ${\bp}_t = m\cdot{\bp}$,
  for ${\bp} = (0,0,1),(0,1,1),(0,0,2)$, have been
  extracted for $m=4,3,2$ respectively and are fitted into
  the dispersion relation:
  $  {{E^2\left(m,{\bp}_t\right)}\over m^2} - ({{\bf c}\cdot {\bp}_t \over m})^2 =
     {{E^2\left(m,{\bf 0}\right)}\over m^2} \ \ $, 
  where $\ \ {\bp}_t = m\cdot {\bp}$
  returning $|{\bf c}|=1.015(32)$,
  which confirms that these many hadron systems are 
  describable in special relativity \mbox{(thermal states would have
  a different dependence).}

  \subsection{One species from single source}
\begin{figure}
      \centering
             \includegraphics[width=6.5cm]{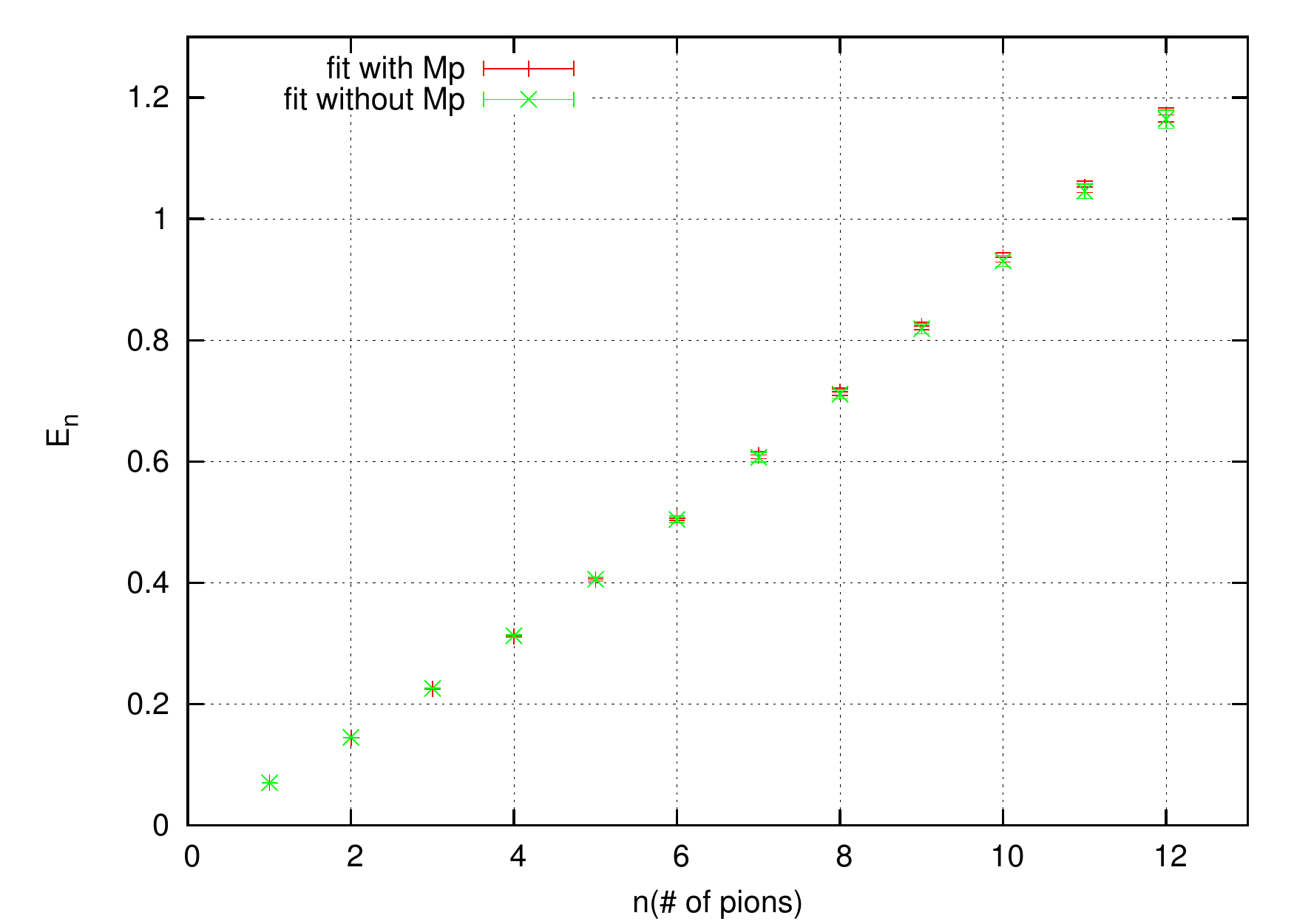}
             \includegraphics[width=7.0cm,angle=0]{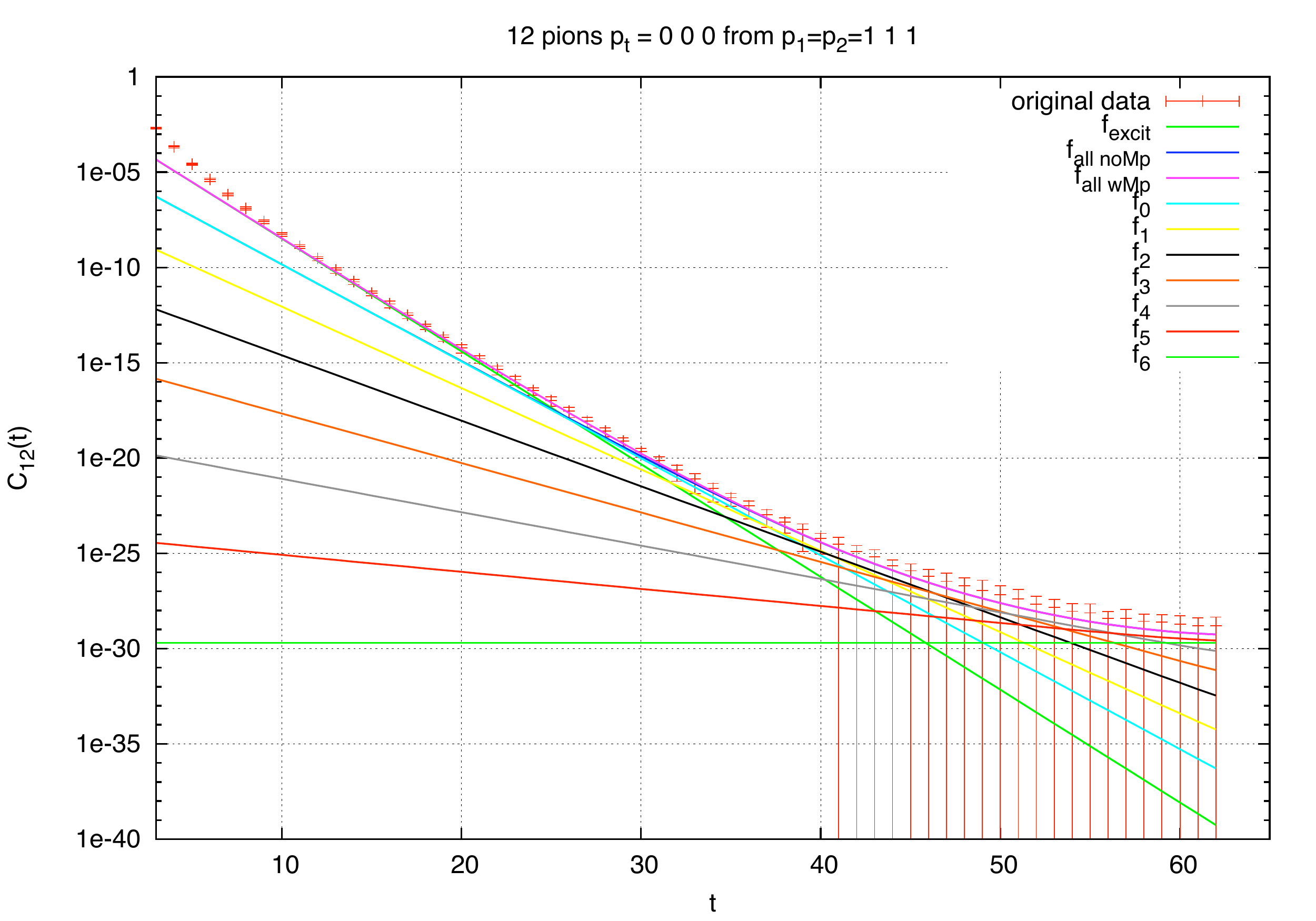}
       \caption{
              The left panel shows energies of
               a rest system of $n$-$\pi^+$($E_{n\pi^+}$) extracted
               from both methods and
                the right panel shows
                $C_{12\pi^+}(t)$, which is decomposed into contributions from 
                excited state, the ground state and all thermal states.}
       \label{fig:energy_000_from_111}
    \end{figure}

    For notational convenience, in the following
    ${\bp}_1^1$ is denoted as ${\bp}_1$, ${\bp}_2^1$ as ${\bp}_2$,
    ${\bp}_1^2$ as ${\bp}_3$ and ${\bp}_2^2$ as ${\bp}_4$.
    Azimuthal symmetry ensures many combinations of ${\bp}_1$, ${\bp}_2$
    be separate measurements of the same physics,
    which provide
    more statistics. As $E_{n\pi^+}$ extracted
    from different choices of ${\bp}_1$ and ${\bp}_2$ agree with each other
    within errors, we choose ${\bp}_1={\bp}_2=(1,1,1)$ for further
    discussions. Extracting $E_{n\pi^+}$ by fitting to functions
    with and without one excited state in addition
    to the thermal states discussed above
    from different time intervals
    produces consistent results, \mbox{as is shown in
    Fig.~\ref{fig:energy_000_from_111}.}

 Decomposing $C_{n\pi^+}(t)$ into different contributions
 gives insight into how much each
 state in Eq.~\ref{eqn:corr_expected} contributes. 
 In Fig.~\ref{fig:energy_000_from_111}, $C_{12\pi^+}(t)$ is shown. In this figure, the green line is
 from the first excited state, the blue one is the ground state and other lines
 are thermal states. It is remarkable that the zero temperature ground state is not dominant in
 any region of this correlator. Even for $C_{12\pi^+}(t)$ the ground 
 state is strongly contaminated by both thermal states and excited
 states, and it becomes difficult to extract the ground state energy
 from $C_{n\pi^+}(t)$, $n > 12$, with the temporal extent of these configurations.

  \subsection{One species from two sources}
\begin{figure}
   \centering
   \subfigure{\includegraphics[width=7.0cm]{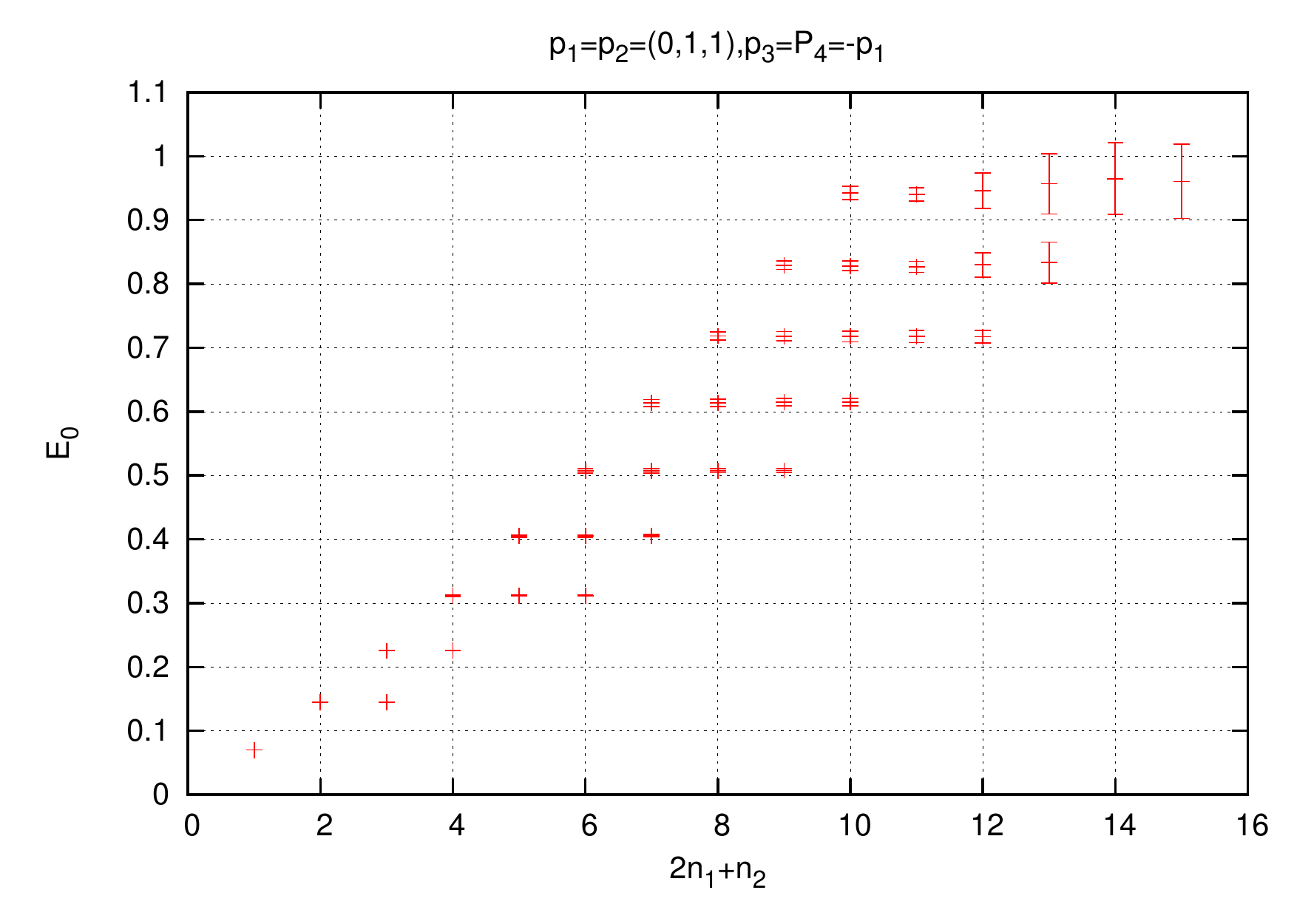}}
   \hspace{0.5cm}
   \subfigure{\includegraphics[width=6.5cm]{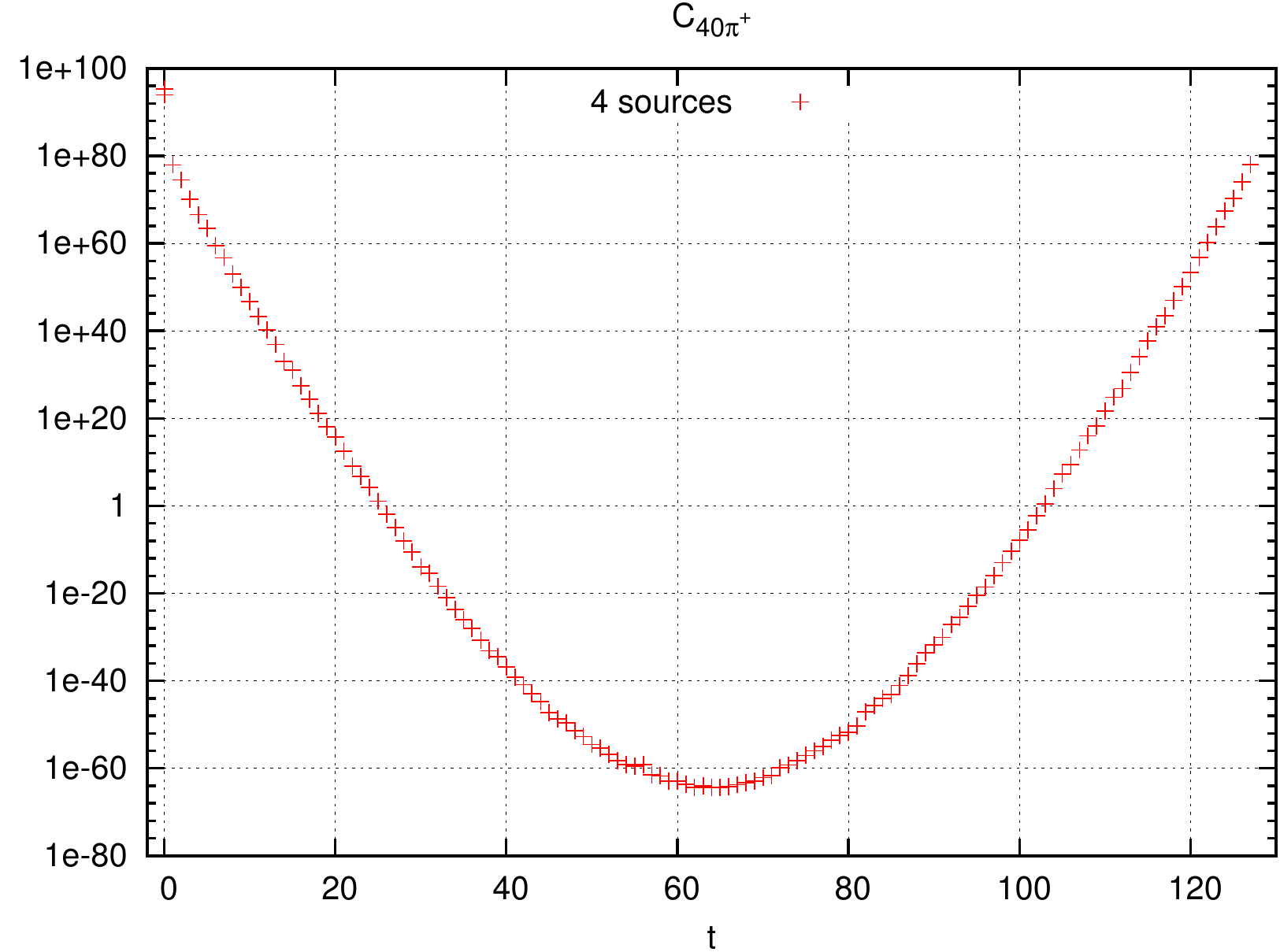}}
   \caption{Left panel shows $E_{n_1\pi^+,n_2\pi^+}$ extracted from $C_{n_1\pi^+,n_2\pi^+}$
     plotted against \mbox{$x=2n_1+n_2$}.
     Statistical error and systematic error are added up in quadrature.
     Right panel is $C_{40\pi^+}(t)$ calculated from the new method.}
   \label{fig:two_sources_all_conf_E0}
 \end{figure}

 By choosing ${\bp}_1={\bp}_2$, ${\bp}_3={\bp_4}$ but ${\bp}_1 \ne {\bp}_3$
 correlation functions of systems having up to $24 \pi^+$ have been computed with
 the same recursion relation, but $E_{n\pi^+}$ is hard to extract for $n>12$.
 As there are more ways to construct a $n$-$\pi^+$ system,
 the recursion relation forces us to calculate all $Q_{n_1\pi^+,n_2\pi^+}$
 for all pairs $n_1,n_2 = 1 \cdots 12$,
 before getting to $Q_{12\pi^+,12\pi^+}$, which costs $O(100)$ times more
 than the one source case. Similarly studying system
 of $36$ $\pi^+$'s requires a third source,
 and becomes $O(100)$ times more expensive again.
 In order to overcome this difficulty, we have developed
 a new method which is  much faster than 
 the recursion relation method and can easily study
 systems of at least 72 mesons~\cite{inpreparation}. 
 This new method is based on the fact that 
 the ground state energies extracted from all $C_{n_1,n_2}(t)$
 for fixed $n_1+n_2$
 are the same
 as they correspond to the same spectrum but with 
 different overlaps, shown in Fig.~\ref{fig:two_sources_all_conf_E0}. 
 
  For systems containing large number of $\pi^+$'s, thermal 
  states dominate the correlation function at later time slices,
  while excited states dominate earlier time slices, and extracting 
  the ground state energy becomes extremely difficulty and systematic
  error is hard to control. In order to get better ground state
  signals we have also studied the system on $20^3 \times 256$ lattices
  (with other parameters unaltered)
  using the new method mentioned above.

 \subsection{Isospin chemical potential ($\mu_I$) and pressure}
\begin{figure}
      \centering
             \subfigure{\includegraphics[width=7.0cm]{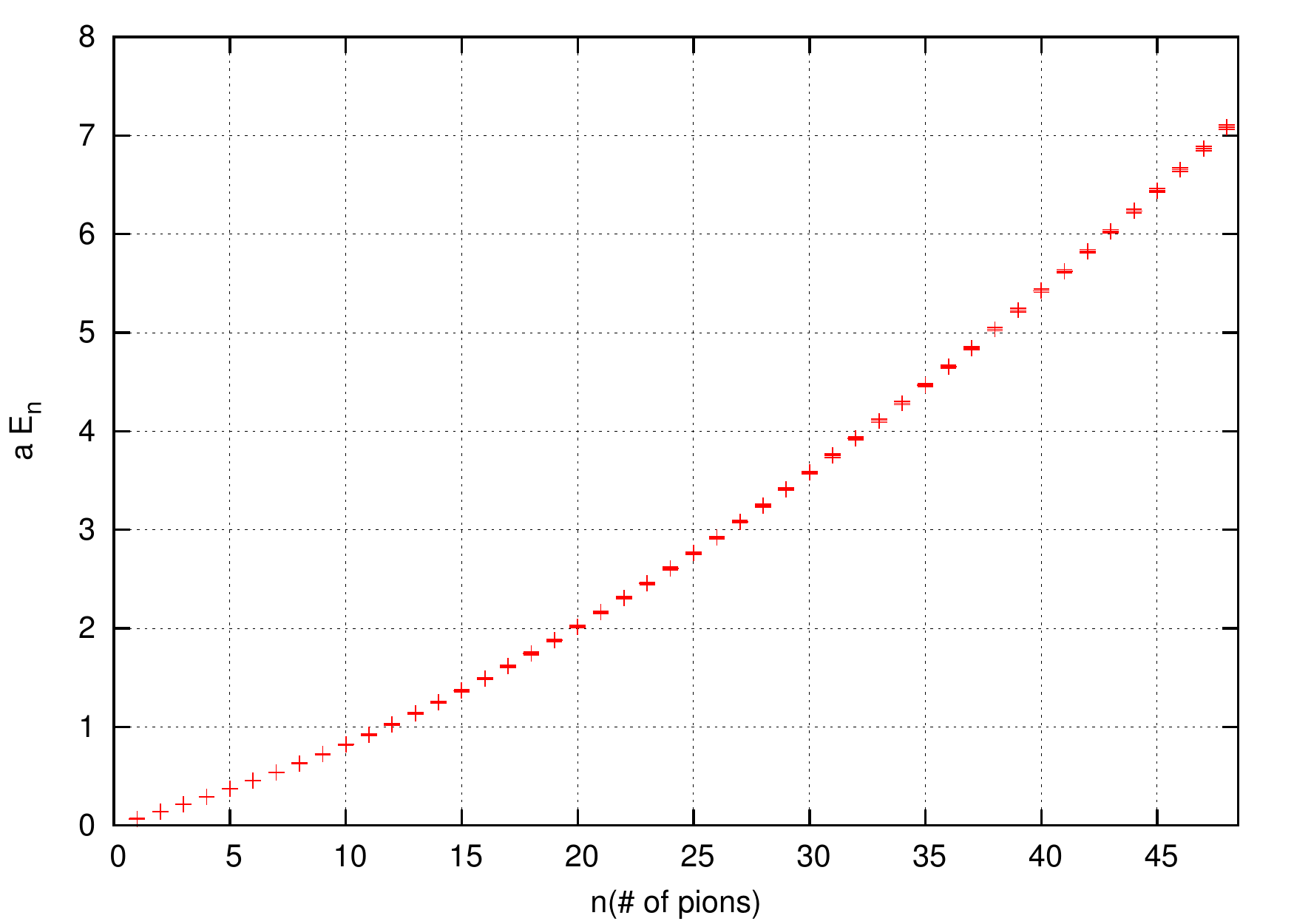}}
             \hspace{0.5cm}
             \subfigure{\includegraphics[width=7.0cm]{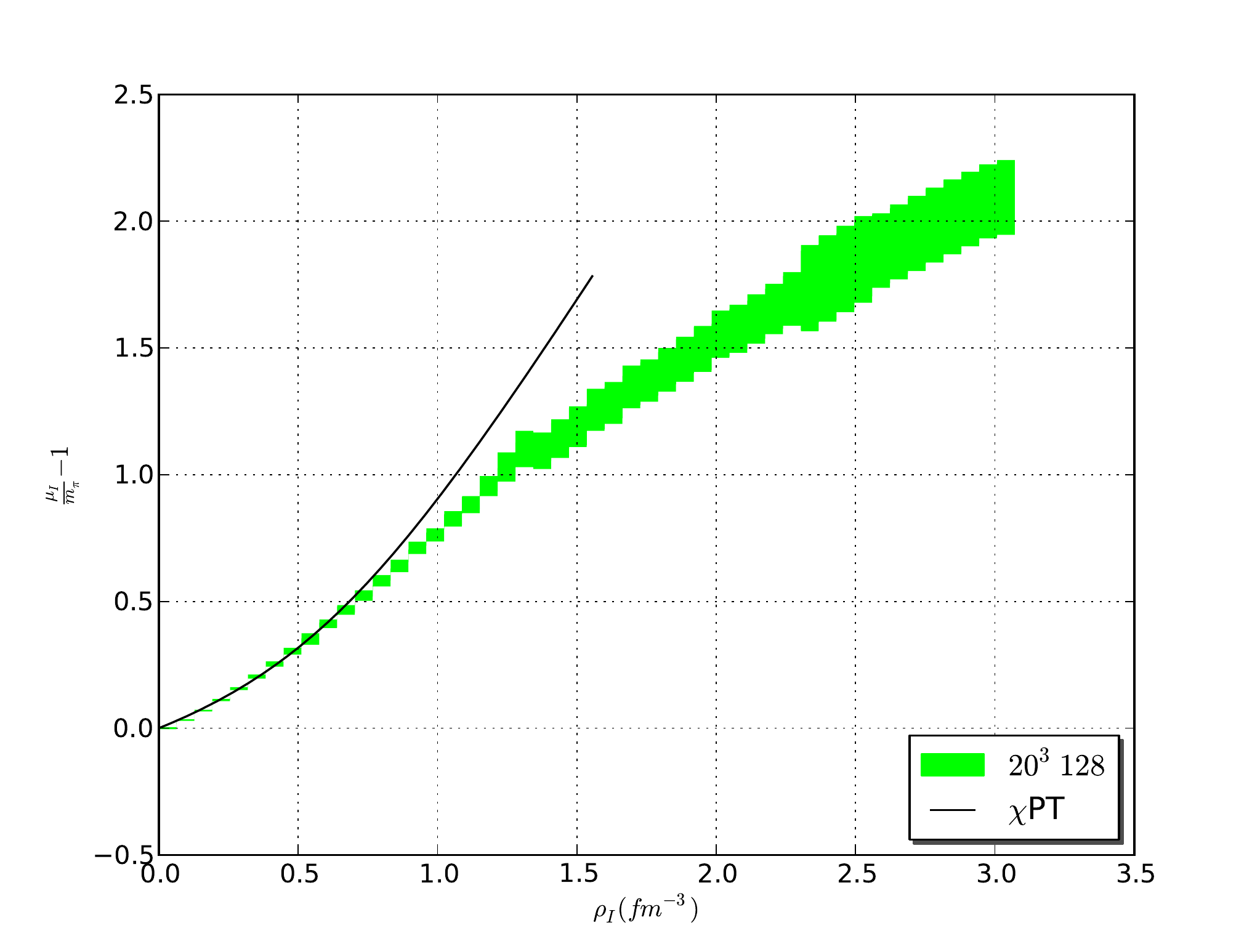}}
       \caption{
              The left panel shows energies of
               a rest system of $n$-$\pi^+$($E_{n\pi^+}$) extracted
               from $20^3 \times 256$ lattice, and
                the right panel shows the isospin chemical potential
	        as a functions of isospin density. Lattice space $a=0.125fm$.
                 The solid black line is from $\chi$PT \cite{Son:Stephanov}
                }
       \label{fig:energy_chemical_from_256}
    \end{figure}

   The isospin chemical potential, the $\mu_I$, is defined as
   ${\mu}_I\left( n \right ) = {{dE}\over{dn}}$,
  which is approximated by using a backward finite difference
    on the lattice. As the systematic errors of ground
   state energies extracted from the $T=128$ lattices are
   large for systems with large number of 
   $\pi^+$'s, we do not extract $\mu_I$
   from these lattices. The $20^3 \times 256$ lattices
   give much better ground state signals, having
   an extremely long plateau in the effective mass
   plot, and a one exponential fit is enough to 
   get the ground state energy. In Fig.~\ref{fig:energy_chemical_from_256},
   we show the ground state energies and isospin 
   chemical potentials from the \mbox{$20^3 \times 256$}
   lattice calculated with the newly constructed method
   mentioned above. $\mu_{\ I}$ is consistent with predictions from
   $\chi$PT \cite{Son:Stephanov} at small density and
   starts to deviate from them for larger densities. 

\begin{figure}
     \centering
      \includegraphics[width=10.0cm]{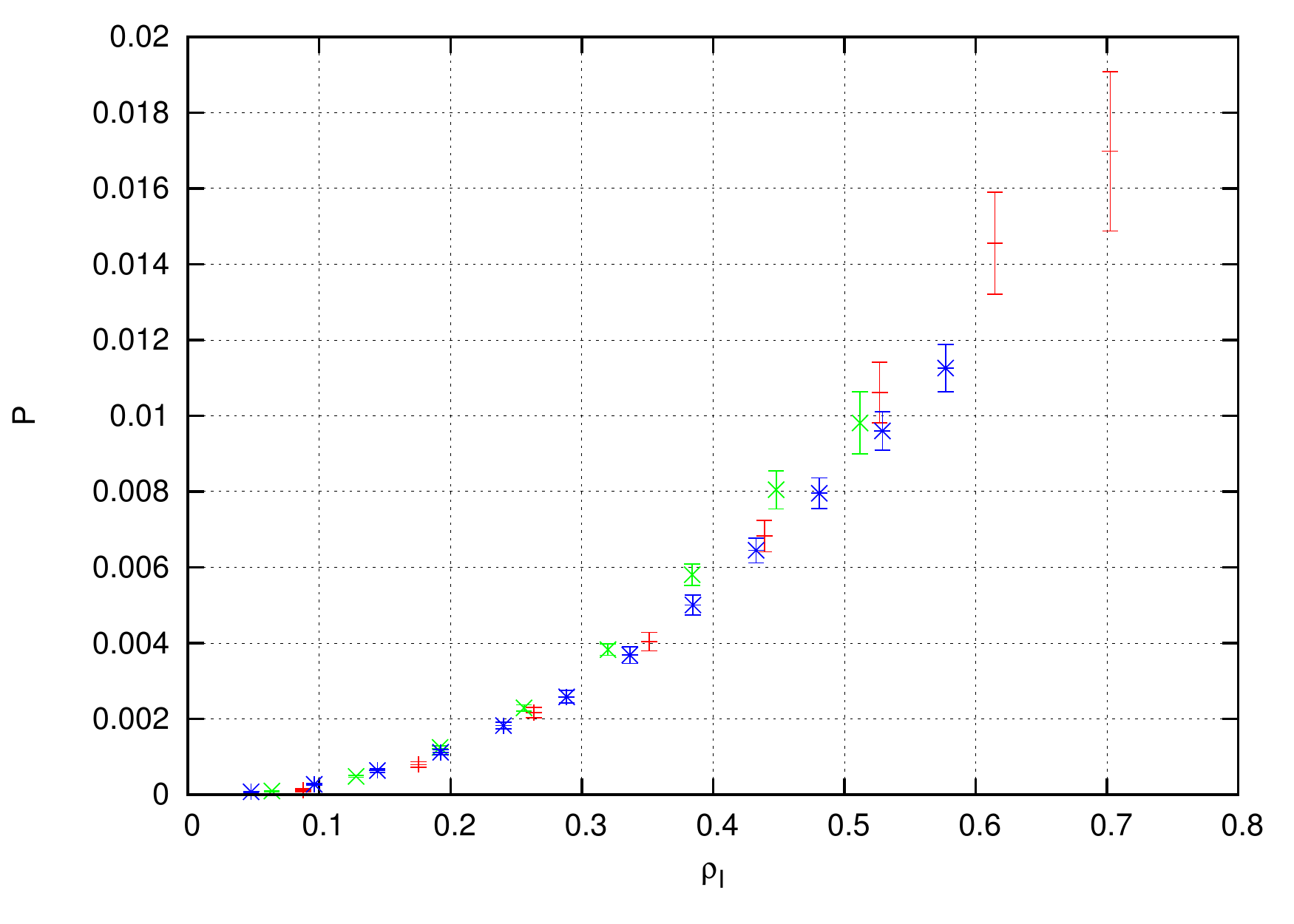}
       \caption{The pressure is computed from energy differences for fixed 
         $n$ on different lattice sizes. Red points are from $16^3$-$20^3$ lattices,
          green points are from $16^3$-$24^3$ lattices, and blue points are from
          $20^3$-$24^3$ lattices. }
      \label{fig:pressure}
     \end{figure}

   Since calculations have been done on several lattices sizes
   $\{16^3, 20^3, 24^3\} \times 128$,
    the pressure ($P$) can be derived as function of density by utilizing
    the discrete version of $P = \frac{dE}{dV}$ for fixed number of $\pi^+$.
    One definition is the following:
    \begin{eqnarray}
      P(\rho_I) = \frac{E_{n}^{L_i}-E_{n}^{L_j}}{(L_i\cdot a)^3-(L_j\cdot a)^3} \nonumber \\
      \rho_I = \frac{n}{((L_i+L_j) \cdot a/2)^3}
     \end{eqnarray}
    where $L_i,L_j = 16,20,24$ and $a=0.125fm$.
    Although we can not
    extract $E_{n\pi^+}$ reliably for large $n$ on these
   ensambles, the pressure can still be studied as
    a function of isospin density from $E_{n\pi^+}$ for small $n$. 
    The pressure is plotted as a function of isospin density in Fig.~\ref{fig:pressure}.

\section{Conclusion}

  We have calculated correlation functions of systems up to $24 \pi^+$s
  in momentum space by applying 
  the recursion relation of Ref.~\cite{Detmold:Savage},
  verified the dispersion relation, and extracted ground state
  energies of $n$-$\pi^+$ systems. Because of the contamination
  from thermal states in later time slices and excited states in
  earlier time slices, it is extremely difficult to see the ground
  state of systems having more than $12 \pi^+$'s for $20^3 \times 128$ anisotropic
  lattices ($a_t \sim 0.04$) at this pion mass.
  In order 
  to reduce contamination from thermal states, we
 also use $20^3 \times 256$ lattices.
   Correlation functions from these larger temporal extent lattices
  give clear ground state signal, and a one exponential fit is sufficient
  enough to reliably extract the ground state energy even \mbox{for $72$ mesons.}
 
  Although the recursion relation requires much less time than 
  direct contractions, it becomes $O(100)$ times more expensive with
  an additional source. Studying 
  $3$ sources system becomes quite expensive
  and studying $4$ sources system becomes impractical.
  In order to overcome this difficulty, we have developed a 
  new method. This new method is far more efficient than
  the recursion relation method, 
  allowing us to study systems of up to $72$ mesons~\cite{inpreparation}.

 \section{Acknowledgement}
  We thank K.~Orginos for valuable discussions.
  The work of WD and ZS is supported in part by JSA, LLC under
  DOE contract No. DE-AC05-06OR-23177 and by the 
  Jeffress Memorial Trust, J-968. WD is
  supported by DOE OJI Award DE-SC000-1784 and DOE grant
  DE-FG02-04ER41302.

\end{document}